\documentclass[12pt]{article}

\usepackage{amsmath,amssymb, amsthm}

\title{Analysis of the role of von Neumann's  projection postulate in the  canonical scheme of 
quantum teleportation and main quantum algorithms}

\author{Andrei Khrennikov\\
School of Mathematics and Systems Engineering\\
International Center of Mathematical Modeling\\ in Physics and Cognitive Sciences\\
V\"axj\"o University, S-35195, Sweden}

\begin{document}

\maketitle

\abstract{ Modern development of quantum technologies based on 
quantum information theory  
stimulated analysis of proposed 
computational, cryptographic and teleportational schemes from the viewpoint of quantum foundations.
It is evident that not all mathematical calculations performed in complex Hilbert space can be directly 
realized in physical space.   Recently by analyzing the original EPR paper we  
  found that they argument was  based on the misuse 
of the von Neumann's projection postulate. Opposite to von Neumann, 
Einstein, Podolsky and Rosen  (EPR) applied   this postulate to observables represented by 
operators with degenerate spectra. It was completely forbidden by von Neumann's 
axiomatics of QM. It is impossible to repeat the EPR considerations in the von Neumann's framework.
 In this note we analyze quantum teleportation 
by taking into account von Neumann's projection postulate.  
Our analysis shows that so called quantum teleportation 
is impossible in von Neumann's framework. On the other hand, our analysis implies that 
 the main quantum algorithms are totally consistent with von Neumann's projection postulate. } 

\section{Introduction}

As a consequence of tremendous development of theoretical basis of quantum information theory, 
quantum technologies became an established domain of experimental research (in particular in laser physics) 
 which is 
directed toward realization of  market-oriented projects in future . Such a situation stimulated analysis of proposed 
computational, cryptographic and delectation schemes from the viewpoint of quantum foundations.
It is evident that {\it not all mathematical calculations performed in complex Hilbert space can be directly 
realized in physical space,}  cf. e.g. \cite{Vol1} -- \cite{HPL2}. 
In particular, quantum information theory (if its is not considered as a purely mathematical formalism) should 
be coupled to quantum measurement theory. The most extensive analysis of  foundations of  
this theory was performed by von Neumann \cite{VN}.  His book became really the Bible of the 
so called Copenhagen interpretation of QM.\footnote{We recall that by this interpretation 
any state of an {\it individual  physical system} is described by
 a wave function $\psi.$ This interpretation was created by  Bohr, Heisenberg, Pauli, 
von Neumann, Fock, Landau. It is commonly used in quantum experimental research. It is important for our further
considerations to point out that: "The state of a system after measurement is determined by the von Neumann
projection postulate." This interpretation is typically confronted with so called statistical (or ensemble) interpretation.
By the latter a wave function $\psi$  is not an attribute of a single physical system (e.g. electron).  
A wave function  $\psi$ (as well as a density matrix $\rho)$  describes an ensemble of identically 
prepared physical system. Here the projection postulate determines not the state of a system 
(after the corresponding measurement), but the probability distribution of an ensemble of (output-)systems.
This interpretation is due to Einstein, Podolsky, Rosen, De Broglie, Schr\"odinger, Bohm,..., Ballentine, 
De Baere,  Manko, Khrennikov,... 
Its modern version is often called the V\"axj\"o interpretation of QM, see \cite{VXU}, \cite{VXU1}: contextual statistical 
realist interpretation of QM.  The latter interpretation combines
the ensemble interpretation of quantum state (not only mixed, but even pure!) with Bohr's principle of 
complementarity.  We remark that, although the the ensemble interpretation of quantum state is widely used in 
theoretical research, e.g. quantum tomography \cite{M1}--\cite{M3}, 
it is still not common for quantum experimental physics.}

 Therefore by considering different constructions 
which arise in quantum information theory we should check their matching with (in particular)  
von Neumann's theory of quantum measurement. 
One of the most intriguing modern constructions of quantum information theory 
is the quantum teleportation scheme \cite{QT}.  In this note we shall analyze it from the viewpoint of 
the quantum measurement theory. As it might be already expected from our analysis of the EPR paper \cite{QT1},   
 the quantum teleportation scheme does not match  von Neumann's theory of measurement. 
In principle, one can proceed up to the very last step of  the quantum teleportation scheme. 
However, to finalize the quantum teleportation  procedure, Bob should perform a measurement in the 
basis which is unknown for him.

On the other hand, our analysis implies that  the main quantum algorithms are totally consistent with 
von Neumann's projection postulate.

\section{Von Neumann's projection postulate}

Von Neumann performed a very deep analysis of the measurement process described by QM. 
One of the fundamental questions which he studied was determination of the output state after measurement. 
Suppose that an observable $A$ was measured and the value $A=a$ 
was obtained. This measurement was performed for some initially prepared state, for simplicity 
we assume that it was a pure state $|\psi >$.  What is a post-measurement state corresponding to this result?  
The answer  is given by the projection postulate \cite{VN}:

\medskip

Part 1:  If the spectrum of the operator $A$ (we shall use the same symbol for an observable and the corresponding 
operator) is  nondegenerate,  then the post-measurement state is given by the eigenvector corresponding to 
the eigenvalue $a$.   

This part of the projection postulate is well known. But, unfortunately, the second (not less fundamental) 
part of the projection postulate is missing in the majority of books and papers:

\medskip

Part 2:  If the spectrum of the operator $A$  is  degenerate,  then the post-measurement {\it state is not determined}. 
To determine it, a refinement measurement should be performed. 
 
\medskip  
   
 By a refinement measurement von Neumann understood measurement of any observable $C$  
which is compatible with $A$ and represented by the operator with nondegenerate spectrum  
such that results of  the $A$-measurement can be obtained  as a function of results of the $C$-measurement, 
i.e., $A=f(C)$.
    
 The crucial point is that in all experiments with {\bf entangled systems}  $S_1+S_2$ 
partial  measurements  on such systems (e.g. on one of particles)  are described by operators 
with degenerate spectra, see \cite{QT1} for details.  Acting in the tensor product $H_1\otimes H_2$ induces degeneration 
(which was absent in e.g. $S_1$-Hilbert space $H_1).$  Thus such a measurement does not determine state.  
To determine the out-put state, one should perform a refinement measurement. But such a refinement measurement 
is always performed on both particles. The latter fact destroys all dreams about quantum nonlocality and its possible
technological applications.

\section{Analysis (due to von Neumann) of the teleportation scheme}

We shall proceed across the quantum teleportation scheme, see \cite{QT}   or simply the corresponding article
in wikipedia,  and point to applications of the projection postulate. 
There are  Alice (A) and Bob (B), and Alice has a qubit in some arbitrary quantum state $|\psi >$. 
Assume that this quantum state is not known to Alice and she would like to send this state to Bob. 
Suppose Alice has a qubit that she wants to teleport to Bob. This qubit can be written generally as: 
$|\psi >=\alpha |0> + \beta|1>.$

The quantum teleportation scheme requires Alice and Bob to share a maximally entangled state before, 
for instance one of the four Bell states:
$\vert \Phi^+>=\frac{1}{\sqrt{2}}(\vert 0 >_A \otimes \vert 0>_B + \vert 1>_A \otimes \vert 1>_B),
\vert \Phi^->=\frac{1}{\sqrt{2}}(\vert 0 >_A \otimes \vert 0>_B - \vert 1>_A \otimes \vert 1>_B),
\vert \Psi^+>=\frac{1}{\sqrt{2}}(\vert 0 >_A \otimes \vert 1>_B + \vert 1 >_A \otimes \vert 0>_B),
\vert \Psi^->=\frac{1}{\sqrt{2}}(\vert 0 >_A \otimes \vert 1>_B - \vert 1>_A \otimes \vert 0>_B).$
Alice takes one of the particles in the pair, and Bob keeps the other one. We will assume that Alice 
and Bob share the entangled state $\vert\Phi^+>.$ 
So, Alice has two particles (the one she wants to teleport, and $A$, one of the entangled pair), 
and Bob has one particle, $B$. In the total system, the state of these three particles is given by
$$\vert \psi> \otimes \vert \Phi^+>=(\alpha \vert 0> + \beta \vert 1>) \otimes \frac{1}{\sqrt{2}} (\vert 0> \otimes \vert 0> + \vert 1 \otimes \vert 1>)$$
Alice will then make a partial measurement in the Bell basis on the two qubits in her possession. To make the result of her measurement clear, we will rewrite the two qubits of Alice in the Bell basis via the following general identities (these can be easily verified):
$\vert 0 > \otimes \vert 0 >=\frac{1}{\sqrt{2}}(\vert \Phi^+> + \vert \Phi^->),
\vert 0 > \otimes \vert 1> =\frac{1}{\sqrt{2}}(\vert \Psi^+> + \vert \Psi^->),
\vert 1 > \otimes \vert 0 >=\frac{1}{\sqrt{2}}(\vert \Psi^+> - \vert \Psi^->),
\vert 1 > \otimes \vert 1> =\frac{1}{\sqrt{2}}(\vert \Phi^+> - \vert \Phi^->),$
Evidently the results of her (local) measurement is that the three-particle state would collapse 
to one of the following four states (with equal probability of obtaining each):
$\vert \Phi^+> \otimes (\alpha \vert 0> + \beta \vert 1>),
\vert \Phi^-> \otimes (\alpha \vert 0> - \beta \vert 1>),
\vert \Psi^+> \otimes (\beta \vert 0> + \alpha \vert 1>),
\vert \Psi^-> \otimes (-\beta \vert 0> + \alpha \vert 1>).$
The four possible states for Bob's qubit are unitary images of the state to be teleported.
The crucial step,  the local measurement done by Alice on the Bell basis,  is done. 
It is clear how to proceed further. Alice now has complete knowledge of the state of the three particles; 
the result of her Bell measurement tells her which of the four states the system is in. 
She simply has to send her results to Bob through a classical channel.  Two classical bits can communicate 
which of the four results she obtained.
After Bob receives the message from Alice, he will know 
which of the four states his particle is in. Using this information, 
he performs a unitary operation on his particle to transform it to 
the desired state $\alpha\vert 0> + \beta\vert1>:$

If Alice indicates her result is $\vert \Phi^+>$, Bob knows his qubit is already in the desired state and does nothing. 
This amounts to the trivial unitary operation, the identity operator.

If the message indicates $\vert \Phi^->$, Bob would send his qubit through the unitary gate given 
by the Pauli matrix
$\sigma_3= \left[ \begin{array}{ll}
1&0\\
0&-1
\end{array}
 \right ]
 $ 
to recover the state.
If Alice's message corresponds to $\vert \Psi^+>$, Bob applies the gate
$\sigma_1= \left[ \begin{array}{ll}
0&1\\
1&0
\end{array}
 \right ]
 $ 
to his qubit.
Finally, for the remaining case, the appropriate gate is given by
$\sigma_3\sigma_1=i\sigma_2= \left[ \begin{array}{ll}
0&1\\
-1&0
\end{array}
 \right ].
 $ 
Teleportation is therefore achieved.

\medskip

The main problem is that Alice's measurement is represented by a degenerate operator 
in the 3-qubit space. It is nondegenerate with respect to her 2-quibits, but not in the total space. 
Thus the standard conclusion that by obtaining e.g. $A=1$, Alice can be sure that Bob obtained the 
right state $\vert \psi>$, does not match the quantum measurement theory. According to von Neumann,  
to get this state Bob should perform a refinement measurement. To to perform it Bob, should know 
the state $\vert \psi>$. It seems that it is the end of the story about quantum teleportation.
We remark that  quantum teleportation does trivially not applicable if one uses so 
called statistical \cite{BL}  (or ensemble) interpretation of QM.  However, in this paper we used not it, 
but the conventional Copenhagen interpretation due to von Neumann.

A number of people (e.g. Richard Gill and Marcus Appleby)  who commented my reanalysis 
of the role of projection postulate
 in the modern version of the Copenhagen interpretation and especially in quantum information
pointed out that, even if Einstein, Podolsky and Rosen used the von Neumann projection postulate
in the improper way, the EPR-version of the projection postulate is nowadays commonly used.
 I agree with them that, although EPR did not use the real von Neumann postulate,  but 
they used the version of the projection postulate which is nowadays known as  Luders' 
postulate \cite{LUDER}  (by which  it is possible to consider operators with
 degenerate spectra in the same way as nondegenerate), see \cite{QT1} for details.  Here the out-put state is given 
by the projection of the  original state onto the corresponding  eigenspace.  But the price of this use is quantum 
nonlocality which also became canonical in modern QM.  It was completely absent in the original 
Copenhagen interpretation, since if one follows J. von Neumann no trace of quantum nonlocality 
would be found. I recall that so called EPR 
states were studied in details in von Neumann's book, but without EPRs paradoxical consequences: 
either incompleteness or nonlocality. One could not completely exclude the possibility that the original 
von Neumann's analysis of quantum measurements was wrong and that, in spite of his demand for sharp 
distinguishing of measurements for observables with degenerate and nondegenerate spectra, it is always possible 
to apply L\"uders' postulate.  We also remark that conclusion of  an ultimate 
 von Neumann's mathematical analysis of quantum measurement procedures totally coincide with views of Niels 
Bohr, see his reply to Einstein \cite{BR}. In such a case it should be openly pointed out that the interpretation of QM which 
is commonly used in quantum information theory  is not the conventional Copenhagen interpretation of 
Bohr-Heisenberg-von Neumann-Fock-Landau-Pauli, ..., but a new "quantum information interpretation of QM." Opposite 
to the conventional Copenhagen interpretation, its QI-version is based on L\"uders' postulate and hence 
quantum nonlocality. We recall that, although EPR had also used  L\"uders' postulate, they still considered 
nonlocality as totally unphysical. Their output from misuse of the von Neumann projection postulate 
was "naive realism" -- assigning values of two incompatible physical variables to the same physical system.
By the QI-interpretation of QM such a possibility is excluded, because of Bell's theorem.  Although Bell's theorem 
does not provide an ultimate proof of this statement, see e.g. \cite{QTG}  and 
\cite{HPL2} and literature hereby,  I would agree with rejection
of naive realism. However, I point out that, opposite to the QI-interpretation, the conventional Copenhagen 
interpretation could peacefully escape both nonlocality and naive realism via the application 
of the proper version (namely,  von Neumann's one) of the projection postulate.\footnote{Thus it could be done 
without Bell's inequality.} 

\medskip

{\bf Conclusion}. {\it If one proceed in the Copenhagen framework in the proper way,  
the quantum teleportation scheme  would not work.}     

\section{Quantum algorithms and   von Neumann's projection postulate}

\subsection{Deutsch-Jozsa algorithm}

Let us start with the Deutsch-Jozsa algorithm.  Since we are interested only in the final measurement 
(unitary evolutions given by quantum gates are not important for our analysis),  we just write the out-put
state of quantum computation:
\begin{equation}
\label{DJA}
\vert \psi> = \frac{1}{\sqrt{2^n}} \sum_{z=0}^{2^n-1} \sum_{x=0}^{2^n-1} (-1)^{xz + f(x)} \vert z> \otimes 
\Big[ \frac{\vert 0> - \vert 1>}{\sqrt{2}}\Big].
\end{equation}
Then we perform measurement of "what is written"  in the argument register $z.$ We consider the operator
\begin{equation}
\label{OPP}
\widehat{P} = \sum_{z=0}^{2^n-1} z  \vert z>\otimes \vert z>.
\end{equation}
This is the operator with nodegenerate spectrum in the "argument-space". Since the final state $\vert \psi>$
is factorized into the "argument state",   
$$
\frac{1}{\sqrt{2^n}} \sum_{z=0}^{2^n-1} \sum_{x=0}^{2^n-1} (-1)^{xz + f(x)} \vert z>
$$
and the "function state"  $\Big[ \frac{\vert 0> - \vert 1>}{\sqrt{2}}\Big],$  we can forget about the last one.

\subsection{Simon's algorithm}

We now move to Simon's algorithm. Here  the output state after  a cicle of quantum computation is given by  
\begin{equation}
\label{SA}
\vert \psi> = \frac{1}{\sqrt{2^n}} \sum_{k=0}^{2^n-1} \Big(\frac{1}{\sqrt{2^n}} \sum_{j=0}^{2^n-1}
(-1)^{jk}\vert j > \otimes \vert f(k)>\Big).
\end{equation}

Although this state is not factorized, we can proceed as it will be descfribed in the following subsection.

\subsection{Application of von Neumann's quantum formalism to partial measurements}

First, we recall von Neumann's formalization of Born's probability postulate (the probabilistic interpretation 
of the wave function):

\medskip

(PI) {\it The probability that in the state $\psi$ the quantity
with  operator $\widehat R$ take on values from an  interval $\Delta_1$ is
\begin{equation}
\label{PB}
P_\psi (R\in \Delta_1) = ||E (\Delta_1) \psi ||^2,
\end{equation}
where $E (\lambda)$ is the  resolution of
the identity belonging to $\widehat R.$}

\medskip

We point out that  $\widehat R$ need not have a nondegenerate spectrum!
Thus the PI-postulate and the projection postulate are not identical in their structures. 
One might say that PI is closer to L\"uders'  postulate (which is simply von Neumann's postulate 
without taking into account the structure of spectrum).

Let $H_1$ and $H_2$ be two complex finite dimensional Hilbert
spaces, $\dim H_i \geq 2.$ Let $\widehat a_1: H_1 \to H_1$  be a 
self-adjoint operator. The
Hilbert space $H_i$ represents (quantum) states of the system
$s_i, i= 1, 2.$ The operator $\widehat a_1$ represents an
observable $a_1$ corresponding to measurements on $s_1.$
The composite system $s=(s_1, s_2)$ is described by the tensor
product space $H=H_1 \otimes H_2.$

Suppose that $\widehat a_1$ has purely discrete nondegenerate spectrum:  $\alpha_j, j=1,..., N= 
\dim H_1.$ 

We remark that measurement of $a_1$ on $s_1$ can be considered as a measurement  
on $s=(s_1, s_2).$  It is represented by the operator 
 $\widehat A_1= \widehat a_1 \otimes I.$ Its spectrum coincides with spectrum of  
$\widehat a_1.$

Suppose  that $s=(s_1, s_2)$ is described by a state $\psi \in H.$
We performed the $a_1$-measurement.  Let $a_1=\alpha_j.$ 

\medskip

PROB). On the one hand, it is a measurement   on $s=(s_1, s_2).$  It is described by 
$\widehat A_1. $ Thus by PI the probability to get   $a_1=\alpha_j$ is given by 
$$
P(a_1= \alpha_j) = \Vert E_{\alpha_j}\otimes I \psi\Vert^2.$$
Thus the probability of the result of a partical measurement (i.e. on one system) 
is given by the PI-postulate for the state space of the composite system $s.$

\medskip

PROJ). On the other hand, the same measurement can be considered as simply a measurement 
on $s_1$ (if systems are isolated at the moment of measurement!). We recall that in $H_1$
the      spectrum of $\widehat a_1$ is nondegenerate. Hence, by the von Neumann's projection 
postulate the resulting state in $H_1$ would always be the eigenstate $\vert \alpha_j>$ of 
$\widehat a_1$ -- independently of the initial state of $s_1.$ The latter independence from the initial state
is important, because by determining the state $\vert \psi>$ of the composite system $s,$ we do not (in general) determine 
states of subsystems. Thus, in spite of this difficulty, we are able to determine the resulting state in $H_1$ 
after  a partial measurement. 

\medskip

We now apply von Neumann's formalism to the state (\ref{SA}) which is produced at the 
end of a cycle of Simon's algorithm and the operator (\ref{OPP}) which plays the role of 
$ \widehat a_1$ in previous considerations.  By PROB we shall get the probability 
of  the result $z;$ by PROJ we really get a state $\vert j>$  which is orthogonal to the vector $\vert x>$ determining 
period.

In the same way we can consider Grover's algorithm and Shor's algorithm.   

\medskip
This paper was written during my visit to Danish Technical University (Copenhagen)  which was supported by 
Informatics and Mathematical Modelling grant of this university.    Introductory 
lectures on quantum information which I gave to students and teachers of the department of 
Computer Science and Engineering stimulated my analysis of quantum information schemes. I would like 
to thank Paul Fischer for hospitality and discussions on possibility of practical realization 
 of  quantum information schemes from the point of view of (classical)  computer science.


\begin{thebibliography}{99} 



\bibitem{Vol1} I. V. Volovich,  Quantum Cryptography in space and Bell's theorem,
in : Foundations of Probability and Physics, Ed. A. Khrennikov,
World Sci. 2001, pp.364-372.

\bibitem{Vol2} I.V. Volovich, Towards Quantum Information Theory
in Space and Time, in: Quantum Theory: Reconsideration of
Foundations, Ed. A.Khrennikov, Vaxjo University Press, 2002,
pp.423-440.

\bibitem{Ozhigov} Y. Ozhigov,  Classical collective behavior instead of quantum dynamics,
arXiv:quant-ph/0702237 (2007).
     
\bibitem{Ozhigov1} Y. Ozhigov,  Simulation of quantum dynamics via classical collective behavior
arXiv:quant-ph/0602155 (2006).

\bibitem{Ozhigov2}  Y. Ozhigov,  Amplitude quanta in multiparticle system simulation, 
{\it Russian Microelectronics},  {\bf 35}, 1, 53-65 (2006).

\bibitem{HPL2}  K. Hess and W. Philipp, ``Bell's theorem: critique of proofs with and
without inequalities'', in {\it Foundations of Probability and
Physics}-3,  edited by A. Yu. Khrennikov, AIP Conference
Proceedings Ser.  {\bf 750}, Melville, New York, 2006, pp. 150-157.

\bibitem{VN} J. von Neumann, {\it Mathematical foundations of quantum
mechanics,} Princeton Univ. Press, Princeton, N.J., 1955.

\bibitem{VXU} A.  Khrennikov. Växjö Interpretation of Quantum Mechanics.
http://xxx.lanl.gov/abs/quant-ph/0202107.

\bibitem{VXU1} A. Khrennikov,  On foundations of quantum theory.
In: {\it Quantum Theory: Reconsideration
of Foundations.} Ser. Math. Modelling, {\bf 2},
163-196, V\"axj\"o Univ. Press, V\"axj\"o,  2002.

\bibitem{M1}  V.  I.  Manko,  {\it J. of Russian Laser Research}, 
{\bf 17},  579-584 (1996).

\bibitem{Manko2}   V. I. Manko and E. V. Shchukin, 
{\it J. Russian Laser Research}, {\bf 22}, 545-560 (2001).

\bibitem{Manko1} M. A. Manko, V. I. Manko, R. V. Mendes,  
{\it J. Russian Laser Research}, {\bf 27}, 507-532.

\bibitem{M2}  S.  De Nicola,  R.  Fedele, M. A. Man'ko and V. I. Man'ko,
Quantum tomography, wave packets, and solitons. 	{\it J. of Russian Laser Research,}
{\bf 25}, 1071-2836,  2004.

\bibitem{M3}  O.  V. Manko and V. I. Manko,  {\it J. Russian Laser Research}, {\bf 25}, 477-492 (2004).

\bibitem{QT}  C.  H.  Bennett, G. Brassard, C. Crépeau, R. Jozsa, A. Peres, W. K. Wootters, 
Teleporting an Unknown Quantum State via Dual Classical and Einstein-Podolsky-Rosen Channels, 
{\it Phys. Rev. Lett.},  {\bf 70},  1895-1899 (1993). 

\bibitem{QT1} A. Khrennikov,   On the problem of completeness of QM: 
von Neumann against Einstein, Podolsky, and Rosen.  arXiv:0805.1511.
 

\bibitem{LUDER}  G. L\"uders,  {\it Ann. Phys., Lpz},  {\bf 8}, 322 (1951).

\bibitem{BL}  L. E. Ballentine, {\it Quantum mechanics,} Englewood Cliffs, New
Jersey, 1989.

\bibitem{BR} N. Bohr,  Phys. Rev. 48, 696-700  (1935).
        
\bibitem{QTG}  A. Khrennikov,  
Analysis of explicit and implicit assumptions in theorems of J. von Neumann and J. Bell.
{\it J. of Russian Laser Research}, {\bf 28}, 244-254 (2007).

  
\end{thebibliography}
\end{document}